\documentstyle[12pt,mont98_desy,amssymb,epsfig]{article}

\def\ra{\rightarrow}

\def\be{\begin{equation}}
\def\ee{\end{equation}}
\def\bea{\begin{eqnarray}}
\def\eea{\end{eqnarray}}
\def\ra{\vec{r}} \def\raq{\vec{r}_{op}}

\def\pa{\vec{p}} \def\pb{\vec{p}_{op}} 
\def\pc{\vec{\pi}} \def\pd{\vec{\pi}_{op}}

 \def\Haq{J_{op}}  
  \def\Hbq{H_{op}}

 \def\Phb{\phi_{op}}

 \def\Bb{\vec{B}_{op}}

\def\Ea{\vec{E}} \def\Eb{\vec{E}_{op}}

 \def\Dab{\vec{\alpha}_{op}}
 \def\DGb{\vec{\gamma}_{op}}
 \def\Dgab{\gamma_{op,4}} 
\def\Dgbb{\gamma_{op,5}} 
\def\DSb{\vec{\Sigma}_{op}} 
\def\Ps{\vec{\sigma}} \def\Psb{\vec{\sigma}_{op}}
\begin{document}

\title{V.~~THE SEMICLASSICAL FOLDY--WOUTHUYSEN TRANSFORMATION AND THE
 DERIVATION 
OF THE BLOCH EQUATION FOR SPIN--1/2 POLARISED BEAMS USING WIGNER FUNCTIONS
\footnote{Updated version of a talk presented
at the 15th ICFA Advanced Beam Dynamics Workshop: ``Quantum
Aspects of Beam Physics'', Monterey, California, U.S.A., January 1998.
Also in DESY Report 98--096, September 1998.}
}
\author{K. HEINEMANN and {\underline {D.P. BARBER}}}

\address{Deutsches Elektronen--Synchrotron, DESY, \\
 22603 Hamburg, Germany. \\E-mail: heineman@mail.desy.de,
mpybar@mail.desy.de}

\maketitle
\abstracts{ A semiclassical Foldy--Wouthuysen transformation of the
Dirac equation is used to obtain the radiationless Bloch equation
for the polarisation density.}
\section{ Introduction and Motivation}
A complete picture of spin polarisation in accelerators and storage rings,
either with or 
without synchrotron radiation, can only be obtained on the basis of
evolution equations for combined spin--orbit distributions.  See Article I.
Moreover, if we are concerned with the effects of
radiation, its simulation by  classical white noise and damping 
does not suffice for all situations. For example we cannot obtain the 
Sokolov--Ternov effect by that means.
In fact to include all the subtleties of radiation, a  quantum
mechanical approach is needed and then obtaining the `complete picture'
implies that we must begin by finding the equation of motion for the
spin--orbit density operator in the presence of radiation.
To ensure some level of transparency and trackability one begins by ignoring
direct and indirect
inter--particle effects so that at the classical level the beam would be
described by a single-particle density depending on the six 
orbital phase space variables, the spin variables and on time as in
statistical mechanics in `$\mu$-space'. 

In the  single-particle approximation, only
positive energy two-component spin-orbit wave functions are needed.
The appropriate quantum Hamiltonian is provided by a
${\bf Foldy-Wouthuysen}$ (FW) transform \cite{fw50} of the Dirac Hamiltonian 
and in order to get explicit results for time-dependent electromagnetic fields
one has to use perturbation theory. Since we are interested in high energy
behaviour in storage rings 
we do {\bf semiclassical} perturbation theory, where the expansion
parameter is Planck's constant, not 1/m etc. 
Before launching into the full blown calculation of the effects of radiation
one should first obtain the transformed Hamiltonian for motion due to the 
Lorentz forces in the fields of the storage ring and then the corresponding 
equations of motion 
for the spin and orbital parts of the density operator. The 
evolution equations for the resulting classical distributions should then be 
derived. These are the tasks of this paper. Radiation will be considered 
elsewhere.

The required Hamiltonian was already stated at first order in $\hbar$ by 
Derbenev and Kondratenko \cite{dk73} as the starting point for their radiation
calculations, but no construction was given. 
In this paper we show how to obtain the transformed Hamiltonian 
to second order in $\hbar$.
Owing to space limitations we will be very
brief but full details can be found elsewhere \cite{kh98};
see also \cite{Cas54,Men54,Ford,Pla66}.
%Then the Weyl transform of the density
%operator (the Wigner function) leads to the above mentioned density and its 
%evolution equation

%The particles are not only subjected to the Lorentz force of the
%external field but to various other effects (synchrotron radiation, spin
%effects). Hence to get the evolution equation for the density one has to 
%include not only the Liouville terms due to the Lorentz force
%  but various other terms.
%Mickey Mousing with single particles relies on assumptions
%\item The normal semi--classical limit ($O(\hbar)$) is 
%sufficient (in retrospect) but do $O(\hbar^2)$ to begin.
%
\section{The Dirac equation}
The underlying Dirac-equation is
$i\hbar({\partial\psi}/{\partial t}) = \Hbq\psi$
where:
\begin{eqnarray}
&&  \Hbq :=  c(\Dab\cdot\pd)
             +m c^2\Dgab +e\Phb \nonumber \\ && \qquad
-\underbrace{
  \frac{e(g-2)\hbar}{4 m c}(
\Dgab(\DSb\cdot\Bb)
      +(\DGb\cdot\Eb))}_{\rm Pauli\;term} \; . \qquad
\nonumber
\end{eqnarray}
We use the matrix representation of \cite{FG}.
%
%\begin{eqnarray}
%&& \Dga:=  \left( \begin{array}{cccc}
%       1 & 0 & 0 & 0\\
%       0 & 1 & 0 & 0\\
%       0 & 0 & -1 & 0\\
%       0 & 0 & 0 & -1
%                \end{array} \right) \; , \qquad
%  \Dg:= i\left( \begin{array}{cc}
%       0 & -\Ps \\
%      \Ps &   0
%                \end{array} \right) \; , \qquad
%  \Dgb = \left( \begin{array}{cccc}
%       0 & 0 & -1 & 0\\
%       0 & 0 & 0 & -1\\
%       -1 & 0 & 0 & 0\\
%      0 & -1 & 0 & 0
%                \end{array} \right) \; , 
%\nonumber \\
%  i\cdot\Dga\cdot\Dgb = \left( \begin{array}{cccc}
%       0 & 0 & -i & 0\\
%       0 & 0 & 0 & -i\\
%       i & 0 & 0 & 0\\
%       0 & i & 0 & 0
%                \end{array} \right) \; , \nonumber\\
% \Da  := \left( \begin{array}{cc}
%       0 & \Ps \\
%      \Ps &   0
%                \end{array} \right) \; , \qquad
%\DS  := \left( \begin{array}{cc}
%       \Ps & 0 \\
%       0  & \Ps
%                \end{array} \right) \; .
%\nonumber
%&&\Dga\cdot\DS  = \left( \begin{array}{cc}
%       \Ps & 0 \\
%       0 & -\Ps 
%                \end{array} \right) \; 
%\label{eq:op22}
%\end{eqnarray}
%
\section{The perturbation theory}
A unitary operator $U_{op}$ which performs the FW transformation  
(a `unitary FW operator') transforms a wave function $\psi$ in the
Dirac representation into the corresponding wave function $U_{op}\psi$ in
the FW representation. In the FW representation the Dirac equation reads as
%
%\begin{eqnarray}
%&&
$i\hbar  \frac{\partial\psi}{\partial t} = \tilde{H}_{op} \psi $
%\; ,
%\nonumber
%\end{eqnarray}
%
where
\begin{eqnarray}
&& \tilde{H}_{op}:=
 U_{op}\Hbq U_{op}^{-1}
 +i\hbar\frac{\partial U_{op}}{\partial t} 
 U_{op}^{-1}  
\label{eq:300}
\end{eqnarray}
and the electron wave functions, i.e. the `positive energy wave functions' 
have only upper components which we denote by $\chi_1,\chi_2$. 
Then in the Dirac representation the positive energy wave functions read as
%
%\begin{eqnarray}
% && 
$\psi = U_{op}^{-1} 
(\chi_1, \chi_2, 0, 0) $.
%\left( \begin{array}{c}
%       \chi_1 \\
%       \chi_2 \\
%        0     \\
%        0
% \end{array} \right)$.
% \; .
%\nonumber
%\end{eqnarray}
% 
Thus in this representation the `positive energy projection operator'
$\Lambda_{op}^{{}_{(+)}}$ reads as
\begin{eqnarray}
&& \Lambda_{op}^{{}_{(+)}} :=
 U_{op}^{-1}\frac{1}{2} (1+\Dgab) U_{op} \; .
\label{eq:10}
\end{eqnarray}
By Eq.~(\ref{eq:10}) the FW transformation (and the FW Hamiltonian)
is not unique because $U_{op}$ can be
replaced by any unitary operator $V_{op}$, such that
$U_{op} V_{op}^{-1}$ is even  
(`essential uniqueness' of the FW transformation).
Our task is now to develop a perturbation theory to construct a
$U_{op}$. Then at
$k$-th order ($k\geq 0$) in $\hbar$,  $U_{op}$ is approximated
by an operator $U_{op,k}$. Starting with zeroth order, to satisfy
Eq.~(\ref{eq:10}), we must choose $U_{op,0}$ such that
\begin{eqnarray}
&& \Lambda_{op}^{{}_{(+)}} \stackrel{0}{=}
U_{op,0}^{-1}\frac{1}{2} (1+\Dgab) U_{op,0} \; .
\label{eq:302}
\end{eqnarray}
Because for semiclassical perturbation theory we have\cite{kh98}:
\begin{eqnarray}
&& \Lambda_{op}^{{}_{(+)}} \stackrel{0}{=}
\frac{1}{2}\biggl(1+ 
 \frac{c}{\Haq}(\Dab\cdot\pd)
             +\frac{m c^2}{\Haq}\Dgab \biggr) \; ,
\label{eq:301}
\end{eqnarray}
we can choose \cite{kh98}:
\begin{eqnarray}
&&{U}_{op,0} :=
\exp\biggl(-\frac{1}{2}\Dgab\Dgbb  
\arctan(\frac{1}{m c}(\DSb\cdot\pd))\biggr)  \; .
\nonumber
\end{eqnarray}
In k-th order Eq.~(\ref{eq:300}) reads as
\begin{eqnarray}
&& \tilde{H}_{op,k}:=
 U_{op,k}\Hbq U_{op,k}^{-1}
 +i\hbar\frac{\partial U_{op,k}}{\partial t} 
 U_{op,k}^{-1}  \; ,
\nonumber
\end{eqnarray}
where $\tilde{H}_{op,k}$ denotes the $k$-th order approximation of the FW
Hamiltonian. Having fixed $U_{op,0}$ the remaining task is to choose
the $U_{op,k}$ for $k\geq 1$ such that $\tilde{H}_{op,k}$ is even
in $k$-th order. That this procedure leads to a unique energy projection 
operator  can be shown 
\cite{kh98} by using
a method due to Pursey \cite{pursey}. 
For \\ 
$k\geq 1$ we compute the $U_{op,k}$ recursively in the form \\   
$ U_{op,k} := \tilde{U}_{op,k} \tilde{U}_{op,k-1}\cdots 
\tilde{U}_{op,1}{U}_{op,0}$, where:
\begin{eqnarray}
&& \tilde{U}_{op,j} := 
\exp\biggl(\frac{1}{4}\lbrack\frac{1}{\Haq}\Dgab
 \; , \; {\cal O}_{op,j-1}\rbrack\biggr) \; , \qquad (j\geq 1) \; ,
\nonumber
\end{eqnarray}
with $ \Haq := \sqrt{m^2 c^4 + c^2(\pd\cdot\pd)}$
 and where ${\cal O}_{op,j}$ is  the odd part of the operator
$\tilde{H}_{op,j}$, so that 
${\cal O}_{op,j}$ anticommutes with $\Dgab$ and 
$\tilde{H}_{op,j}-{\cal O}_{op,j}$ commutes with  $\Dgab$.
Thus ${\cal O}_{op,j}$ is the odd part obtained after the $(j+1)$-th step.
\par In first order one gets 
$ \tilde{H}_{op,1}
 \stackrel{1}{=} \tilde{H}_{op}^{dk}$ \cite{kh98}
where:
\begin{eqnarray}
&& \tilde{H}_{op}^{dk} :=
\frac{1}{2}\Dgab\Haq +\frac{e}{2}\Phb  
-\frac{e\hbar}{4m c}(\frac{mc^2}{\Haq}+
\frac{g-2}{2})\Dgab(\DSb\cdot\Bb)
\nonumber\\&&\qquad
+\frac{e\hbar c(g-2)}{8m}
\frac{1}{\Haq(\Haq+mc^2)} 
\Dgab(\pd\cdot\Bb)(\DSb\cdot\pd)
\nonumber\\&&\qquad
+\frac{e\hbar}{4 m}
\biggl(\frac{g-2}{2}\frac{1}{\Haq} 
+ \frac{m c^2}{\Haq(\Haq+m c^2)}\biggr)
\biggr(\DSb\cdot(\pd\wedge\Eb)\biggr) + {\rm h.c.} \; .
\label{eq:2}
\end{eqnarray}
In second order, i.e. in $\tilde{H}_{op,2}$,
terms quadratic in $\Eb$ and $\Bb$ and 
gradient terms in  $\Eb$ and $\Bb$ (e.g. the `Darwin term') appear but no spin
terms.
\section{The two-component formalism}
%
%In first order the wave equation for $\chi:=(\chi_1,\chi_2)$
%reads as
%
%\begin{eqnarray}
%&& i\hbar\frac{\partial\chi}{\partial t} = h_{op}^{dk}\;\chi \; ,
%\nonumber
%\end{eqnarray}
%
The two-component Hamiltonian $h_{op}^{dk}$, the `DK-Hamiltonian',
is obtained from
$\tilde{H}_{op}^{dk}$ by the replacements
$\Dgab\rightarrow 1$ and $\DSb \rightarrow \Psb$
so that:
\begin{eqnarray}
&& h_{op}^{dk} :=  h_{op,orb}^{dk}  
+ \frac{\hbar}{2}\cdot(\Psb\cdot
\vec{\Omega}_{op}) \; ,
\label{eq:304}
\end{eqnarray}
where $h_{op,orb}^{dk}  := \Haq + e\Phb$ and:
\begin{eqnarray}
&& \vec{\Omega}_{op} :=
-\frac{e}{2 m c}(\frac{m c^2}{\Haq}+
\frac{g-2}{2})\Bb
+\frac{e c(g-2)}{4 m}
\frac{1}{\Haq(\Haq+m c^2)} 
(\pd\cdot\Bb)\pd
\nonumber\\&&
+\frac{e}{2 m}
\biggl(\frac{g-2}{2\Haq} 
+ \frac{m c^2}{\Haq(\Haq+m c^2)}\biggr)
(\pd\wedge\Eb)
 + {\rm h.c.} \; .
\label{eq:305}
\end{eqnarray}
\section{Nonrelativistic approximation}
Expanding $h_{op}^{dk}$ w.r.t. $1/m$ one gets in first order in $1/m$
(`nonrelativistic limit'):
$  m c^2 + ({1}/{2m})(\pd\cdot\pd)  
  + e \Phb 
- ({e\hbar g}/{4 m c})(\Psb\cdot\Bb)$
which for $g=2$ is the Pauli-Schroedinger Hamiltonian.
\section{The density operator in the two-component formalism}
The density operator $\rho_{op}$ reads as:
\begin{eqnarray}
 && \rho_{op} = \frac{1}{2}~(\rho_{op,orb} + \Psb\cdot\vec{\xi}_{op}) \; , 
\label{eq:3}
\end{eqnarray}
where $\rho_{op,orb}$ and $\vec{\xi}_{op}$ do not contain spin degrees of
freedom. The normalisation of the density operator reads as
$ 1 = Tr \lbrack\rho_{op}\rbrack = Tr \lbrack\rho_{op,orb}\rbrack$.
%
%
%(Note: we don't write:
%$\rho_{op} =\frac{1}{2}\cdot\rho_{op,orb} + 
%\frac{1}{4}(\rho_{op,orb}\cdot{\vec P}_{op} + 
%{\vec P}_{op}\cdot\rho_{op,orb } )\cdot\Psb $ ) \\
%
Expanding the Hamiltonian and the density operator to second order,
the evolution equation for the density operator,  
the `von-Neumann equation', reads in first order as:
\begin{eqnarray}
 && 0 \stackrel{1}{=} -\frac{\partial\rho_{op}}{\partial t} + 
\frac{i}{\hbar}\lbrack \rho_{op} \, , \,
h_{op}^{dk} \rbrack \; .
\nonumber
\end{eqnarray}
Note that the second order parts of the Hamiltonian drop out of 
this von-Neumann equation because they are independent of spin.
By Eq.~(8) the von-Neumann equation is in first order 
equivalent to:
\begin{eqnarray}
 && 0 \stackrel{1}{=} -\frac{\partial\rho_{op,orb}}{\partial t} + 
\frac{i}{\hbar}\lbrack \rho_{op,orb} 
 \, , \, h_{op,orb}^{dk} \rbrack + \frac{i}{2}
\biggl((\vec{\xi}_{op}\cdot\vec{\Omega}_{op}) 
- (\vec{\Omega}_{op}\cdot\vec{\xi}_{op})\biggr) \; , 
\label{eq:4} \\
&& 0 \stackrel{1}{=} -\frac{\partial\vec{\xi}_{op}}{\partial t} + \frac{1}{2}
(\vec{\Omega}_{op}\wedge\vec{\xi}_{op}) -
 \frac{1}{2}(\vec{\xi}_{op}\wedge\vec{\Omega}_{op}) + 
\frac{i}{\hbar}\lbrack \vec{\xi}_{op} 
\, , \, h_{op,orb}^{dk} \rbrack  \nonumber\\&&\qquad
+ \frac{i}{2}
\lbrack \rho_{op,orb} \, , \, \vec{\Omega}_{op} \rbrack \; . 
\label{eq:5}
\end{eqnarray}
The terms proportional to $\vec{\Omega}_{op}$ in  Eq.~(\ref{eq:4})
account for the effect of the SG force on the orbital motion and
the second and third terms on the rhs of Eq.~(\ref{eq:5})
 have the same structure 
as the Thomas-BMT equation. 
\par The terms $h_{op,orb}^{dk}$ and $\vec{\Omega}_{op}$ in the 
von-Neumann equation are  not unique because
the FW Hamiltonian 
%(hence $h_{op}^{dk}$)
depends on the chosen FW 
transformation. However, essential uniqueness
allows the forms in Eqs.~(\ref{eq:304}) and (\ref{eq:305}).
\section{The Weyl transform in the two-component formalism}
The `Weyl transform' allows q-numbers to be represented by c-numbers.
In the two-component formalism an operator $K_{op}$ is represented by
its Weyl transform $K_{wt}$ via:
\begin{eqnarray}
 && K_{wt,\nu\lambda}(\ra,\pa;t):=  Tr\lbrack K_{op}
  {\cal M}_{op,\nu\lambda} 
  \Delta_{op}(\vec r,\vec p)\rbrack 
\; , \qquad (\nu,\lambda=1,2) \; ,
\label{eq:op40}
\end{eqnarray}
where the operator $\Delta_{op}$ is defined by:
\begin{eqnarray}
&& \Delta_{op}(\vec r,\vec p) :=
\frac{1}{8\pi^3\hbar^3}
\int\;d^3 x\; d^3 u \exp\biggl(\frac{i}{\hbar}\lbrack
\vec x\cdot(\vec r-\raq) +
\vec u\cdot(\vec p-\pb)\rbrack\biggr) \; , 
\nonumber
\end{eqnarray}
and where the operators ${\cal M}_{op,11},...$
are defined by:
\begin{eqnarray}
&& ( {\cal M}_{op,\nu\lambda}\;\chi)_\mu = \delta_{\mu\nu}\chi_{\lambda} \; ,
\qquad (\mu,\nu,\lambda=1,2) \; .
\nonumber
\end{eqnarray}
Conversely one has:
\begin{eqnarray}
 && K_{op} = \frac{1}{8\pi^3\hbar^3}\sum_{\nu,\lambda=1}^2\;
  {\cal M}_{op,\nu\lambda} 
\int\;d^3 r\; d^3 p\; K_{wt,\nu\lambda}
  \Delta_{op}   \; .
\label{eq:op58}
\end{eqnarray}
Thus the Weyl transform is a $2\times 2$ matrix valued phase space function;
the $t$-dependence in $K_{wt}$ only occurs for time-dependent operators.
In terms of its Weyl transform  the trace of an operator $K_{op}$ reads as:
\begin{eqnarray}
 && Tr\lbrack K_{op}\rbrack =
\frac{1}{8\pi^3\hbar^3} tr\lbrack
\int\;d^3 r\; d^3 p\; K_{wt}\rbrack \; ,
\label{eq:op43}
\end{eqnarray}
where $tr$ denotes the matrix trace.
\section{The Wigner function}
The chosen normalisation of the density operator $\rho_{op}$
leads to:
\begin{eqnarray}
 && 1 = \frac{1}{8\pi^3\hbar^3} tr\lbrack
\int\;d^3 r\; d^3 p\; \rho_{wt}\rbrack \; ,
\label{eq:6}
\end{eqnarray}
for its Weyl transform $\rho_{wt}$  (see Eq.~(\ref{eq:op43}))
and one calls $(1/8\pi^3\hbar^3)\rho_{wt}$ the `Wigner function'
of that state. In terms of Weyl transforms
the expectation value of an operator $K_{op}$ reads as:
\begin{eqnarray}
 && <K_{op}>
= \frac{1}{8\pi^3\hbar^3} tr\lbrack
\int\; d^3 r \; d^3 p \;\rho_{wt} K_{wt}\rbrack \; .
\label{eq:op103}
\end{eqnarray}
\section{The Wigner-Kirkwood expansion}
Since we are dealing with a beam, which in a 
high energy accelerator occupies a phase space volume many orders of magnitude 
greater then $\hbar^3$, we are very far from dealing with a pure state. Then,  
applying semi-classical perturbation theory to the density operator,
its Weyl transform has the `Wigner-Kirkwood' form:
\begin{eqnarray}
&& \rho_{wt} = 
\underbrace{\rho_0}_{Classical\;part} + 
\underbrace{\hbar\cdot\rho_1+\hbar^2\cdot\rho_2 + ...}
_{Quantum\;corrections} \; ,
\label{eq:8}
\end{eqnarray}
where $\rho_0,\rho_1,\rho_2,...$ are of zeroth order in $\hbar$
\footnote{Note that $\rho_{op}$ has a related expansion and that this was
used in Eqs.~(9) and (10).}.
In reality, `classical distributions' $\rho_0$ do not exist so that 
one has to deal with Eq.~(\ref{eq:8}).
\section{The Weyl transform of the Hamiltonian}
The Weyl transform of the Hamiltonian is
$ h_{wt}^{dk} \stackrel{1}{=} h_{orb}^{dk} + \frac{\hbar}{2}(\Ps\cdot
\vec{\Omega}_{dk})$
where:
\begin{eqnarray}
&& h_{orb}^{dk} :=   J + e\phi  \; , \nonumber\\
&& \vec{\Omega}_{dk} :=
-\frac{e}{m c}(\frac{m c^2}{J}+
\frac{g-2}{2})\vec B
+\frac{c e(g-2)}{2 m}
\frac{1}{J(J+m c^2)}
(\vec\pi\cdot\vec B)\vec\pi \nonumber\\&&\qquad
+\frac{e}{m}
\biggl(\frac{g-2}{2}\frac{1}{J} 
+ \frac{m c^2}{J(J+m c^2)}\biggr)
(\pc\wedge\Ea) \; ,
\nonumber
\end{eqnarray}
and where
%
%\begin{eqnarray}
%&&
$ \vec\pi := \vec p - \frac{e}{c}\vec A$ and
%\; , \qquad
$J := \sqrt{m^2 c^4+c^2(\vec\pi\cdot\vec\pi)}$.
% \; .
%\nonumber
%\end{eqnarray}
%
%
%\begin{eqnarray}
%&& \Delta h_{orb} =
% - \frac{\hbar^2}{2\cdot J}\cdot (J \odot J) \nonumber\\&&\qquad
%-\frac{e^2\cdot\hbar^2\cdot c^3}{8}\cdot\biggl(
%\frac{2\cdot J^2-m^2\cdot c^4}
%{J^4\cdot(J+m\cdot c^2)^2} 
%+ \frac{g-2}{J^3\cdot(J+m\cdot c^2)}\biggr)\cdot
%\pc\cdot(\Ba\wedge\Ea)
%\nonumber\\&&\qquad
% +\frac{e\cdot\hbar^2\cdot c^4}{8}\cdot\biggl(
%\frac{2\cdot J^2+2\cdot m\cdot c^2\cdot J +
%m^2\cdot c^4}{J^4\cdot(J+m\cdot c^2)^2}\nonumber\\&&\qquad
%+\frac{g-2}{m\cdot c^2}\cdot\frac{1}{J^2\cdot(J+m\cdot c^2)}\biggr)
%\cdot\pc\cdot
%((\pc\cdot\vec\nabla) \Ea) \nonumber\\&&\qquad
%-\frac{e\cdot\hbar^2\cdot c^2}{4}\cdot\biggl(
%\frac{1}{J\cdot(J+m\cdot c^2)}
%+\frac{g-2}{2\cdot m\cdot c^2\cdot J}\biggr)
%\cdot\vec\nabla\cdot\Ea 
%%\nonumber\\&&\qquad
%
%
%-\frac{\hbar^2\cdot e^2\cdot c^2}{8\cdot J^3}\cdot
%\Ba\cdot\Ba \nonumber\\&&\qquad
% + \frac{e^2\cdot(g-2)\cdot\hbar^2}{8\cdot m^2}\cdot\biggl(
% m\cdot c^2\cdot\frac{2\cdot J+m\cdot c^2}{J^3\cdot(J+m\cdot c^2)^2}
% + \frac{g-2}{4\cdot J^3}\biggr)\cdot
%(\pc\cdot\Ba)\cdot(\pc\cdot\Ba)
% \nonumber\\&&\qquad
% +\frac{e\cdot\hbar^2\cdot(g-2)\cdot c}{8\cdot m}\cdot
% \frac{1}{J\cdot(J+m\cdot c^2)}\cdot
%\pc\cdot(\vec\nabla\wedge\Ba) \nonumber\\&&\qquad
%-\frac{e^2\cdot\hbar^2}{2\cdot m^2}\cdot\biggl(
%\frac{m^2\cdot c^4}{4}\cdot\frac{1}{J^5}
%-\frac{(g-2)^2}{16}\cdot\frac{1}{J^3}\biggr)\cdot
%(\pc\cdot\Ea)\cdot(\pc\cdot\Ea) \nonumber\\&&\qquad
%+\frac{e^2\cdot\hbar^2\cdot c^2}{8}\cdot\frac{1}{J^3}\cdot
%(1+\frac{g-2}{2})^2\cdot
%\Ea\cdot\Ea \; ,
%\nonumber
%\end{eqnarray}
%
%
%
%
\section{The Weyl transform of the von-Neumann equation}  
The Weyl transform of the von-Neumann equation is an evolution equation
for $\rho_{wt}$. In particular  from 
Eqs.~(\ref{eq:4}) and (\ref{eq:5}) and in  first order using section 7,
one gets:
\begin{eqnarray}
 && 0 \stackrel{1}{=} 
\underbrace{
-\frac{\partial\rho_{wt,orb}}{\partial t} +
 \lbrace 
 h_{orb}^{dk} \, , \,  \rho_{wt,orb} \rbrace  
}_{\rm classical\;Liouville\;terms}
+\underbrace{
 \frac{\hbar}{2}
 \lbrace 
\Omega_{dk,j} \, ,\, \xi_{wt,j} 
\rbrace 
}_{\rm SG\;term}
\; ,  \label{eq:11} \\
 && 0 \stackrel{1}{=} \underbrace{
-\frac{\partial\vec{\xi}_{wt}}{\partial t} 
+ \lbrace 
h_{orb}^{dk} \, ,\, \vec{\xi}_{wt} 
\rbrace 
+ \vec{\Omega}_{dk}\wedge\vec{\xi}_{wt} 
}_{\rm Thomas-BMT\;terms} \;
+\underbrace{
\frac{\hbar}{2}
\lbrace 
\vec{\Omega}_{dk} \, ,\,
\rho_{wt,orb} 
 \rbrace 
}_{\rm SG\;backreaction\;term} \;  ,
\label{eq:12}
\end{eqnarray}
where  $\lbrace \;, \; \rbrace$ is the usual Poisson bracket w.r.t.
$\vec r$ and $\vec p$ and where repeated indices are summed over.
As in Eqs.~(\ref{eq:4}) and (\ref{eq:5})
the second order parts of the Hamiltonian drop out.
The vector ${\vec \xi}_{wt}$  is the {\bf polarisation density} 
and Eq.~(18) is its {\bf Bloch equation} in first order \cite{kh98}.
Note that these equations are not restricted to dipole and quadrupole 
magnetic fields. It is the $\rho_{wt}$ and ${\vec \xi}_{wt}$ which serve 
as the `classical distribution functions' which we have been seeking.

%
%In a purely classical treatment (see Article I)
%$\vec{\xi}_{wt}$ would be replaced by the classical
%{\bf `polarisation density'} 
%$\vec{\xi}_{cl} := \frac{1}{2}
%\rho_{cl,orb} \vec P$
%so that the $\rho_0$ in Eq. (\ref{eq:8}) reads as
%$ \rho_{0} = \rho_{cl,orb} + \frac{1}{2}
%(\Ps\cdot\vec{P})$
%and one has:
%
%\begin{eqnarray}
% && 0 \stackrel{0}{=} 
%-\frac{\partial\rho_{cl,orb}}{\partial t} +
% \lbrace 
% h_{orb}^{dk} \, , \,  \rho_{cl,orb} \rbrace  
%\; , \qquad
% 0 \stackrel{0}{=} 
%-\frac{\partial\vec{P}}{\partial t} 
%+ \lbrace 
%h_{orb}^{dk} \, ,\, \vec{P}
%\rbrace 
%+ \vec{\Omega}_{dk}\wedge\vec{P} \; .
%\nonumber
%\end{eqnarray}
%We call this  equation for the polarisation density a `Bloch equation'.  
%(Note that in Article I the
%symbol $w$ is used for the phase space density instead of $\rho$.)

These equations are easily transformed from cartesian coordinates to 
`machine coordinates' since with the Weyl transform 
one only has to deal with c-numbers instead of q-numbers \cite{bhr1}.
After transforming to machine coordinates 
the zeroth order limits of the transformed Eqs.~(\ref{eq:11}) and (\ref{eq:12})
correspond to the classical
Eqs.~(17) and (40) in Article I.

For FW transformations of the Dirac equation where time has been replaced by
the longitudinal coordinate in a paraxial approximation see \cite{jag}.
\section{Radiation}
Now that the radiationless case is on a firm basis, one can include
radiation. In a classical treatment of radiation effects one gets 
the Fokker-Planck and Bloch equations of Eqs.~(24) and (39) in Article I.
%\begin{eqnarray}
%&&\frac{\partial\rho_{cl,orb}} {\partial t}   =
%       L_{{}_{FP,orb}} \; \rho_{cl,orb} \; ,
%\nonumber
%\end{eqnarray}
%%
%which implies the Bloch equation (see Article I):
%
%\begin{eqnarray}
%&&\frac{\partial\vec{\xi}_{cl}} {\partial t}   =
%       L_{{}_{FP,orb}} \; \vec{\xi}_{cl} +
%\vec{\Omega}_{dk} \wedge \vec{\xi}_{cl} \; .
%\nonumber
%\end{eqnarray}
%%
To include the Sokolov-Ternov effect one needs
a full quantum treatment \cite{dk75}.  
\section*{Acknowledgments}
We would like to thank R. Jagannathan and S. Khan for encouragement and for
useful exchanges of ideas.


\section*{References}
\begin{thebibliography}{99}
\bibitem{fw50}
L. Foldy and S. Wouthuysen, Phys.Rev., {\bf 78}, 29 (1950).

\bibitem{dk73}
Ya.S. Derbenev and A.M. Kondratenko, Sov.Phys.JETP., {\bf 37}, 968 (1973).

\bibitem{kh98}
K. Heinemann, Thesis  in preparation.

\bibitem{Cas54}
K.M. Case, Phys.Rev., {\bf 95}, 1323 (1954).
%
\bibitem{Men54}
H. Mendlowitz, Ph.D. thesis, The University of Michigan (1954).

\bibitem{Ford}
G. Ford, University of Michigan, unpublished notes.

\bibitem{Pla66}
E. Plahte, Supp. Nuovo Cim., {\bf 4}, 246 (1966).

\bibitem{FG}
D.M. Fradkin, R.H. Good, Rev.Mod.Phys., {\bf 33}, 343 (1961).

%\bibitem{roman}
%P. Roman, ``Advanced Quantum Theory'', Addison--Wesley (1965).

\bibitem{pursey}
D.L. Pursey, Nucl.Phys., {\bf 8}, 595 (1958).

\bibitem{bhr1}
D.P. Barber, K. Heinemann and G. Ripken, Z.f.Physik, {\bf C64}, 117--167
 (1994).

\bibitem{jag}
The reader is directed to the articles by R. Jagannathan and S. Khan 
in these Proceedings and references therein.

\bibitem{dk75}
Ya.S. Derbenev and A.M. Kondratenko, Sov.Phys.Dokl., {\bf 19}, 438 (1975).


\end{thebibliography}
\end{document}